\documentstyle[referee]{mn}

\input{epsf}

\title[TF relation ]
{The Radial Tully-Fisher relation for spiral galaxies I   }
\author[Yegorova \& Salucci]
       {Irina A. Yegorova$^{1}$,  Paolo Salucci$^{1}$\\
        $^{1}$SISSA - International School for Advanced Studies,
           via Beirut 2-4,
           Trieste,
           Italy, e-mail:yegorova@sissa.it\\
}

\date{Accepted ...,
      Received ...,
        in original form ... .}

\begin{document}
\maketitle

\begin{abstract}

We find a new  Tully-Fisher-like relation for spiral galaxies
holding at different galactocentric radii. This Radial Tully-Fisher
(RTF) relation allows us to investigate the distribution of matter
in the optical regions of spiral galaxies. This relation, applied to
three different samples of rotation curves of spiral galaxies,
directly proves that: 1) the rotation velocity of spirals is
a good measure of their gravitational potential and both the RC's
amplitudes  and profiles are well predicted  by galaxy luminosity
2) the existence of a dark component, less concentrated than the
luminous one, and 3) a scaling law, according to which,
inside the disk optical size: $ M_{dark}/M_{lum} =
0.5 (L_B/10^{11}L_{B\odot})^{-0.7}$.

\end{abstract}

\section{Introduction}

In 1977 Tully and Fisher discovered that the maximal rotational
velocity $V_{max}$ of a spiral galaxy, measured by the FWHM of
the neutral hydrogen 21-cm line profile, correlates with the
galaxy luminosity by means of a power law of exponent $a \sim
4$. This equivalently reads as:
\begin{equation}
M= a \log{ V_{max}} + b
\end{equation}
with $M$ the absolute magnitude in some specified band and $b$ a
constant. It was immediately realized that this  relation, hereafter
TF, could serve as   a  powerful  tool to determine the distances of
galaxies \cite{Pierce+Tully-1} and to study their dynamics
\cite{Persic+Salucci-1}. The rotational velocity reflects the
equilibrium configuration of the underlying galaxy gravitational
potential, especially when $V_{max}$ is directly derived from
extended rotation curves. Before proceeding further, let us point
out that spiral galaxies  have a characteristic size scale,
$R_{opt}$, that sets also a characteristic reference velocity
$V(R_{opt})$. $R_D$,  the exponential thin disk length scale, is a
natural choice for such reference radius; in this paper, however, we
adopt for the latter a minimal variant, i.e. a  multiple of this
quantity: $R_{opt} \equiv 3.2 R_D$ (see Persic, Salucci \& Stel
1996, hereafter PSS). (No result here depends on the value of the
multiplicity constant). This choice is motivated by the fact that
$3.2 R_D $, by enclosing $83\%$ of the total light, is a good
measure of the "physical size" of the stellar disk,  and that,  for
many purposes, $V_{opt} \equiv V(R_{opt})$.
\footnote{ For the PS95 sample:
$\log{V_{opt}} = (0.08 \pm 0.01) + (0.97 \pm 0.006)\log{V_{max}}$.}

Let us  stress that some known kinematical quantities are not
suitable reference velocities. For example, the value of
$V_{max}$ for a spiral depends on the extension and on the
spatial resolution of the available RC and, in addition,  it
does not have a  clear physical interpretation, sometimes
coinciding with the outermost available velocity measure, in
other cases with the innermost one.  Also $ V_{last}$, the
velocity at the outermost measured point obviously does not have
a proper physical meaning, in addition some spirals never reach
the, so called, asymptotic flat regime (PSS and Salucci \&
Gentile 2006).

Coming back to the  TF relation, its  physical explanation, still not fully
understood, very likely  involves the argument that  in
self-gravitating rotating disks both the rotation velocity and the
total luminosity are a measure of the same  gravitational mass (e.g.
Strauss \& Willick 1995). Notice that, if this argument is correct,
both $V_{max} $ and $V_{last} $ are just empirical quantities of
different and not immediate physical meaning.

The stars in spiral galaxies are settled thin disks with an exponential
surface mass  distribution
\begin{equation}
\Sigma (R) = \Sigma_0 e^{-R/R_D}\,, \quad   \Sigma_0=  k_1
L^s\,,
\end{equation}
where  $\Sigma_0= (M_d/L) I_0$  is the central surface mass
density, with   $I_0$ the central surface  brightness, in the
first approximation, constant in spirals.  $L $ is the total
luminosity in a specific band,  $k_1$ and $s$ are
constants. Observationally, as a good approximation, we find
that (e.g. PSS):
\begin{equation}
R_D= k_3 L^q\,,
\end{equation}
with $q=1/2$, and $k_3={\rm const}$. It is illustrative to set, for
the time being,  $s=0$, i.e. to assume that the disk mass-to-light
ratio has an universal value. Let us consider the condition of
self-gravity equilibrium for the stellar disk, i.e. the ratio
$E={{GM_D}/(V^2_{opt}R_{opt})}$. By writing:
\begin{equation}
E = k_2 L^t\,,
\end{equation}
where $k_2$ and $t$ are constants, we have that Freeman disks are 
completely self-gravitating for $k_2 \simeq 1.1$, $t=0$. Combination 
of the previous equations  and  the above
assumptions leads to  the well known  relation: luminosity $
\propto$ (velocity)$^4$. Random  departures from  the above
conditions will produce a larger scatter in the TF relation,
while systematic departures  from the assumption made above e.g.
variations  of the stellar population with galaxy luminosity or
violation of the condition of self-gravity, will modify the
slope, zero-point and scatter, possibly in a band-dependent way.
In fact, by relaxing some of the assumptions made above, we have
the more general relationship: $L \propto V_{opt}
^{2/{(+s+q-t)}}$ (here $s$, $q$ and $t$ can be band-dependent)
that can be even more complex and non-linear when the scaling
laws (2), (3), (4) are not just power laws. As a matter of fact,
in several different large samples of galaxies it has been found
that the TF has different slope and scatter in different bands:
$a_I \simeq 10$, $s_I \sim 0.4 {\rm mag} $, while $a_B \simeq
7.7$, $s_B \sim 0.5 {\rm mag} $ (Pierce \& Tully 1992),
(Salucci et al. 1993). Moreover,  a non linearity in the TF is
often found at low rotation velocities \cite{Aaronson-et-al}.

We know that spiral galaxies are disks of stars embedded in (almost)
spherical halos of dark matter and this is crucial for understanding
the physical origin of the TF relation (Persic \& Salucci 1988,
Strauss \& Willick 1995, Rhee 1996). It is well known that the dark
halos paradigm is supported by the (complex) mass modelling of
galactic rotation curves  (e.g. PSS and references therein) and it
implies that disks are not fully self-gravitating. At any radius,
both the dark and luminous components contribute to the (observed)
rotational velocity $V(R)$, with a relative weight that varies both
radially and from galaxy to galaxy.  The resulting model circular
velocity  can be written as a function of the useful radial
coordinate $x \equiv R/R_{opt}$ as \footnote{For simplicity here we
neglect the bulge, we will consider it in section 4.} :
\begin{equation}
V_{model}(x)=(G M_D /R_D)^{1/2}\,[f_d(x)/f_d(1)+\Gamma f_h(x,\alpha)]^{1/2},
\end{equation}
where $f_d(x)$ is the Freeman velocity  disk profile
\begin{equation}
 f_d(x)={1\over {2}}(3.2
 x)^2(I_0(1.6 x)K_0(1.6 x)-I_1(1.6 x)K_1(1.6 x))\,,
\end{equation}
and
  \begin{equation}
  f_h(x,\alpha)=({x^2\over {x^2+\alpha^2}})(1+\alpha^2),
  \end{equation}
$M_d$ is the disk mass, $\Gamma $ is the dark/visible matter
velocity ratio at $R_{opt}$ and $\alpha$ is the halo velocity  core radius in
units of $R_{opt}$. The adopted halo function  $f_h(x,\alpha)$ (see PSS) is the
simplest way to describe the contribution of dark matter halo; in fact, for an
appropriate value of $\alpha$ the DM term in $V_{model}(x)$ describes both the
``empirical'' universal rotation curve halo velocity profile in PSS and, in the
regions under study (by setting $\alpha=1/3$)  the "theoretical" $\Lambda CDM $
$V_{NFW}(x)$ Navarro-Frenk-White halo velocity profile.

%%%%%%%%%%%%%%%%%%%%%%%%%%%%%%%%%%%fig1%%%%%%%%%%%%%%%%%%%%%%%%%%%%%%%%%%%
\begin{figure}
\centering\epsfxsize=12.cm \epsfbox{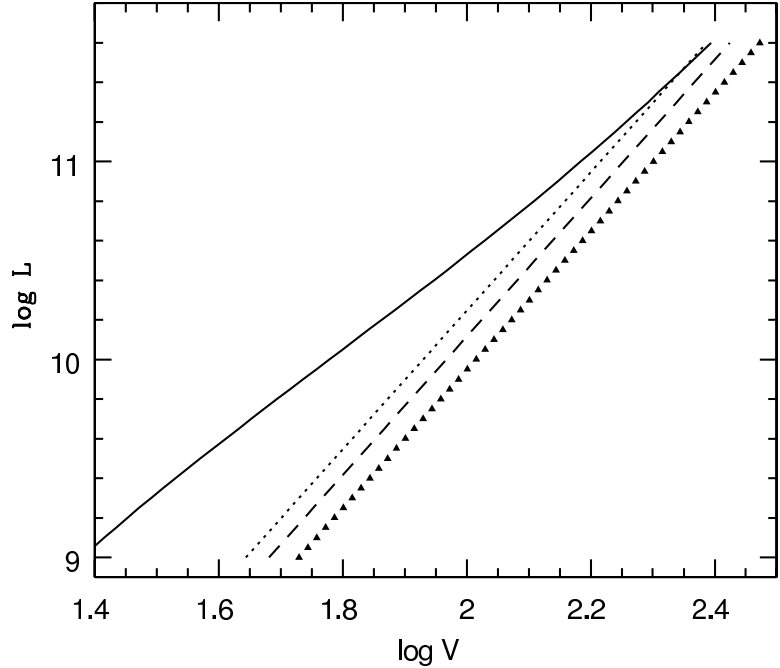}
%\epsfbox{fig1.eps}
\vspace{-5.cm} \caption{The TF relation at $R_{opt}$ ({\it
triangles}); the predicted  relation at $R_{d}\approx(1/3)R_{opt}$
in the case of: no-DM ({\it short dashed line}),
$\Gamma = 1$ NFW halo ({\it dotted line}) and  $\alpha=2$ and
$\Gamma=0.5 (L_B/10^{11} L_{B\odot})$ ({\it solid line}).}
\label{TF1}
\end{figure}
%%%%%%%%%%%%%%%%%%%%%%%%%%%%%%%%%%%fig1%%%%%%%%%%%%%%%%%%%%%%%%%%%%%%%%%%%

Once a spiral mass model is assumed, the well know  TF relation
at $R_{opt}$:
\begin{equation}
 L_B= \left({V_{opt}\over {200 km/s}}\right)^{3.3} 10^{11}\  L_{B\odot},
\end{equation}
(e.g. PSS, Persic \& Salucci 1995
hereafter PS95), may lead through eq.(5)-(7), to
the existence of a family of  similar relationships at radii
$ jR_{opt}$ ($j=0.2 - 1.4$), whose  values for the slopes, scatters
and zero points are all different and all depending on the
characteristics of the mass distribution. In Fig.1 we show this for
the reasonable values of $\Gamma$ and $\alpha$ at.
Of course if $\Gamma$ and $\alpha$ are random functions then no relation 
will emerge.

The aim of this  paper is to  investigate the rotation curves (RCs
elsewhere) of three  samples of spiral galaxies and to extract  from
actual data  the   Radial Tully-Fisher relation (RTF), i.e. a family
of TF-like relations holding at any properly chosen radial distance.
Then, we will use the properties of such relationships to
investigate the mass distribution in these objects.

Let us notice that the strict Universal Rotation Curve paradigm, put
forward in PSS, implies, in addition to   an Universal velocity profile
at any chosen luminosity, also a "TF like" relationship  at any chosen
radius $x$. While in  PSS the  aim was mainly to derive the
above universal profile, in  the  present paper, meant to be
complementary to PSS,   we will head for the latter set of
relationships.

In order to obtain statistically relevant results as free as
possible from biases, observational errors and cosmic variance
effects, we use  three different samples of spiral galaxies with
available rotation curves. These samples mostly contain Sb-Sc spiral
galaxies. The number of early type spirals and dwarfs is very small.
Moreover,  in general, the bulge affects only the first reference
radius. Different investigation will be necessary to
assess the present results in bulge dominated spirals and HI
dominated dwarfs.

The plan of this work is the following. In section two we
describe our data samples and the main steps of the analysis. In
section three we show that TF-like relations hold at specific
radii, and we derive the basic parameters of these relations for
our samples. In the next section we discuss the implications of
the existence of Radial TF relation and we propose a simple mass
model that fits the data. The conclusions are given in section 5.

\section{Data and analysis}

Sample 1 consists of 794 original RCs of PS95 (notice that for most 
of them the limited number of independent measurements and sometimes  
some circular motion make difficult to derive a proper mass model, 
but instead will be possible with the method in section 2,3).

In each RC the data are radially binned on a $0.2 R_{opt}$
scale so that we have 4-7 independent and reliable estimates of
the circular velocity, according to its extension.

Sample 2 from Courteau (1997) consists of 86 RC's (selected from
304 galaxies) and Sample 3 of Vogt (2004) 81 RC's (selected from
329 galaxies). These samples have been built  by selecting from the
original samples only objects  with high quality and high resolution
kinematics yielding reliable determinations of both {\it
amplitudes} and {\it profiles} of the RCs. To ensure this, we have
set the following selection criteria. The RC's must: {\it (a)} extend
out to $\simeq R_{opt}$; {\it (b)} have at least 30 velocity
measurements distributed homogeneously with radius and between the
two arms; and {\it (c)} show no global asymmetries or significant
non-circular motions: the profiles of the approaching and receding
arms must not disagree systematically more than $15\%$ over 1$R_d$
length-scale. The velocity errors are between $1\%-3\%$.

In  each galaxy we measure the distance from its center $R$ in
units of $ R_{opt}$ ($R_{opt} \equiv 3.2 R_D$) and we consider a
number of  radial bins centered at $R_n= (n /5) R_{opt}$  for
the PS95 sample and  at   $ R_n= (n/20) R_{opt} $ for the other
two samples; we take the bin size  $\delta= 0.2 R_{opt}$  for
the PS95 sample and  $\delta= 0.06 R_{opt}$ for the other two
samples. Then we co-add and average  the velocity values that
fall in the bins,  i.e. in the radial ranges $ R_n -\delta/2
\leq R_n  \leq R_n + \delta/2$ and we get the average circular
velocity $V_n$ at  the  chosen reference radii $R_n$. (For the
PS95 sample this was made in the original paper)

A "large" radial bin size has been chosen for the (much bigger) PS95
sample because, by selection, most of its  RCs have a relatively
smaller number of measurements. For the other two samples, that,
exclusively include extended high quality RC and large number of
measurements we decrease the bin size by a factor 3.2.

In short, we will use  two different kind of samples:  sample 1
includes 794 Sb-Sd galaxies  with I magnitudes whose   RCs are
estimated inside large radial bins that smooth out non circular
motions and observational errors; samples 2 and 3 include 167
galaxies with R magnitudes,  whose     RCs of  higher quality are
estimated  inside smaller radial bins providing so a larger number
of independent data per object.

We look for a series of correlations, at the radii $R_n$ between the
absolute magnitude $M$ (in bands indicated below) and  $\log \  V_n
\equiv  \log \  V(R_n)$. Data in the $I$ \cite{Mathewson-et-al-1} and
$r$ \cite{Courteau-1}, \cite{Vogt-et-al-1} bands will allows us to check the
dependence of our results  on the type of stellar populations in spiral
galaxies. Finally let us stress that the uncertainties of photometry are 
about $10\%$ and therefore negligible.

%%%%%%%%%%%%%%%%%%%%%%%%%%%%%%%%%%%fig2%%%%%%%%%%%%%%%%%%%%%%%%%%%%%%%%%%%
\begin{figure}
\centering \epsfxsize=16.cm \epsfbox{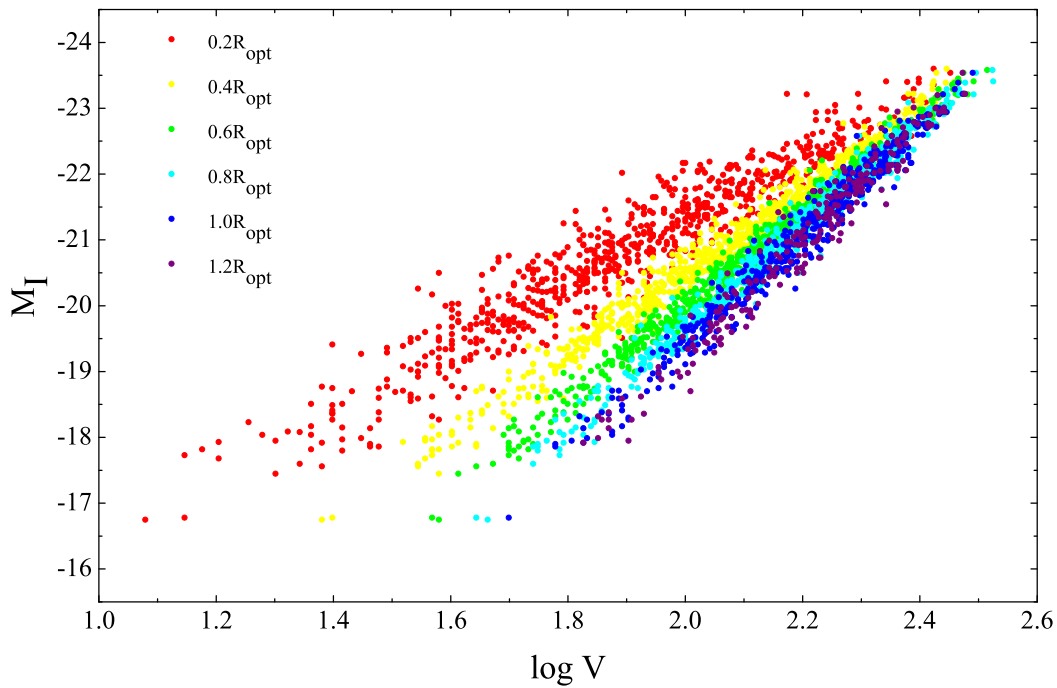}
\vspace{-1.cm}\caption{The Radial Tully-Fisher relations for the
PS95 sample. Each one of the 6 relations is indicated with different
color.} \label{TF2}
\end{figure}
%%%%%%%%%%%%%%%%%%%%%%%%%%%%%%%%%%%fig2%%%%%%%%%%%%%%%%%%%%%%%%%%%%%%%%%%%

\section{The Radial TF relationship}

Given a sample of galaxies of magnitude $M_{band}$ and reliable
rotational curves, the  Radial Tully-Fisher (RTF
further on)  relation is defined, as the  ensemble  of the
fitting relationships:
\begin{equation}
M_{band} = a_n \log V_n + b_n\,,
\end{equation}
with   $a_n$,  $b_n$,  the parameters of the fits and  $R_n$
the radial coordinates at which the relationship is searched.
The latter is defined for all object as a fixed multiple
of the disk length-scale (or equivalently of $R_{opt}$).
Parameters $a_n$, $b_n$ are estimated by the least squares method
(without considering the velocity/magnitude uncertainties).

The existence of the Radial TF relation is clearly seen in Fig. 2
and Fig. 3, where all the TF-like relations for the PS95 sample are
plotted together and identified with  a different color. It is
immediate to realize that they mark an ensemble of linear relations
whose slopes and zero-points vary continuously with the reference
radius $R_n$.

Independent Tully-Fisher like relationships exist in spirals at any
"normalized" radius $R_n $. We confirm this in a very detailed and
quantitative way in the  Figures 11, 12 and in the Tables 1, 2,
3, where very similar results are found for the other two samples.
It is noticeable that the various investigations  lead to the same
consistent picture.

The slope $a_n$ increases monotonically with $R_n $; the scatter
$s_n $ has a minimum at about two disk length-scales,  $ 0.6
R_{opt}$. In the $I$-band the
values of the slopes are about 15$\%$ larger than those in the
$r$-band. This difference, well known also for the standard TF, can
be interpreted as due to the decrease, from the $r$ to the $I$ band,
of the parameter $s$ (see Eq. 2), as effect of a different
importance in the luminosity of the population of recently formed 
stars \cite{Strauss+Willick}.

It is possible to compare the RTFs  in different bands; in
the case of absence of DM, true in the inner regions of spirals
(except in the very luminous galaxies and LSB almost absent in our
sample), for the reasonable values $s_r=0.1$ and $s_I=0 $ the power
law coefficient of the $L_I$ vs velocity relationship
is larger by a factor $(0.5+s_r)/(0.5 +s_I)$ than that of the $L_r$  vs
velocity relationship,  in details by a factor 1.2. This correction allows us
to compare the $a_n$ slopes as a function of $R_n$ for our samples
(see Figure 4). Remarkably, we find that the values of the slopes vary with
$R_n$ according to a specific pattern: $b_n=-2.3-9.9(R_n/R_{opt})+3.9 (R_n/R_{opt})^2$.

%%%%%%%%%%%%%%%%%%%%%%%%%%%%%%%%%%%fig3%%%%%%%%%%%%%%%%%%%%%%%%%%%%%%%%%%%
\begin{figure}
\centering \epsfxsize=17.cm \epsfbox{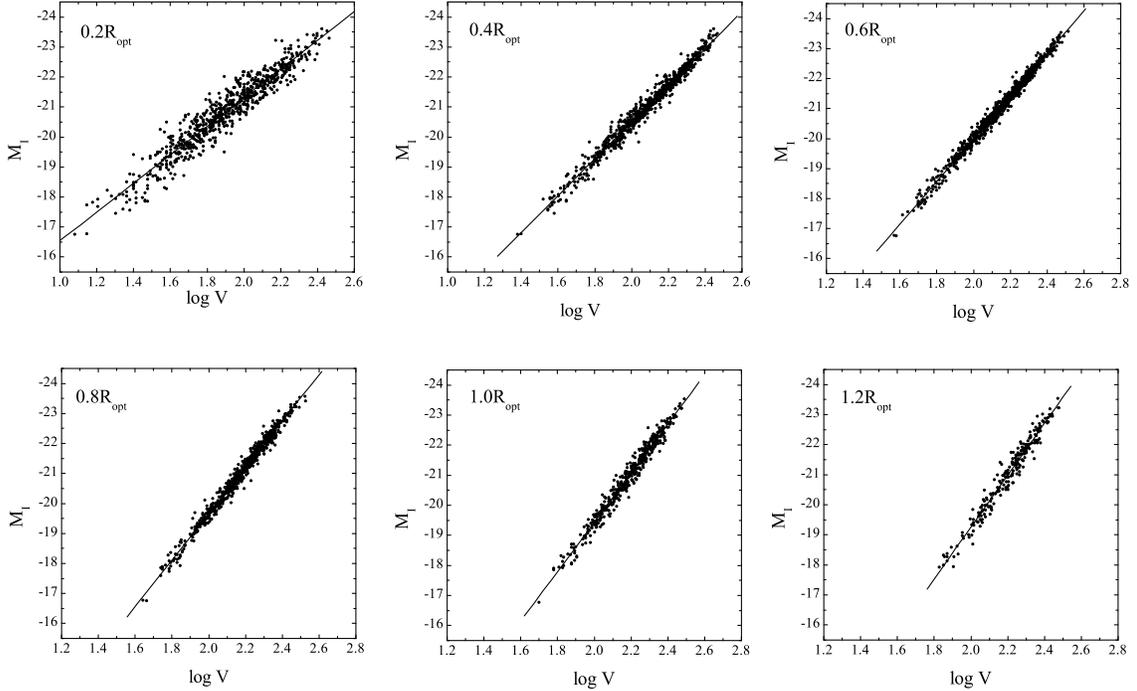} \vspace{-1.cm}
\caption{The Radial Tully-Fisher relation  for the PS95 sample.}
\label{TF3}
\end{figure}
%%%%%%%%%%%%%%%%%%%%%%%%%%%%%%%%%%%fig3%%%%%%%%%%%%%%%%%%%%%%%%%%%%%%%%%%%

%%%%%%%%%%%%%%%%%%%%%%%%%%%%%%%%%%%fig4%%%%%%%%%%%%%%%%%%%%%%%%%%%%%%%%%%%
\begin{figure}
\centering\epsfxsize=12.cm \epsfbox{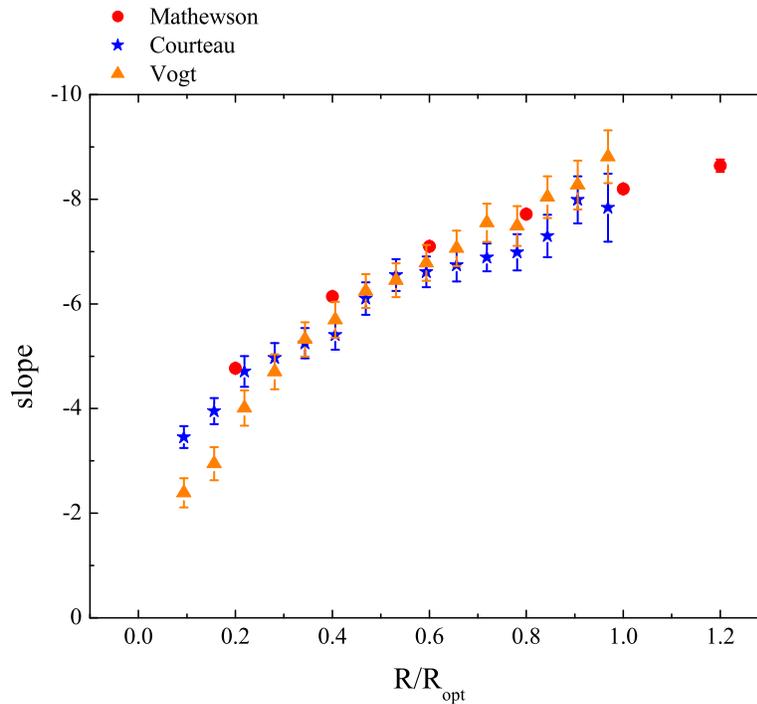}
%\epsfbox{fig1.eps}
\vspace{-1.cm}  \caption{The slope of the Radial Tully-Fisher relation
at different radii for the 3 samples. The slope for the standard
TF is about to -7.5.} \label{TF4}
\end{figure}
%%%%%%%%%%%%%%%%%%%%%%%%%%%%%%%%%%%fig4%%%%%%%%%%%%%%%%%%%%%%%%%%%%%%%%%%%

%%%%%%%%%%%%%%%%%%%%%%%%%%%%%%%%%%%fig5%%%%%%%%%%%%%%%%%%%%%%%%%%%%%%%%%%%
\begin{figure}\vspace{-1.cm}
\centering\epsfxsize=12.cm \epsfbox{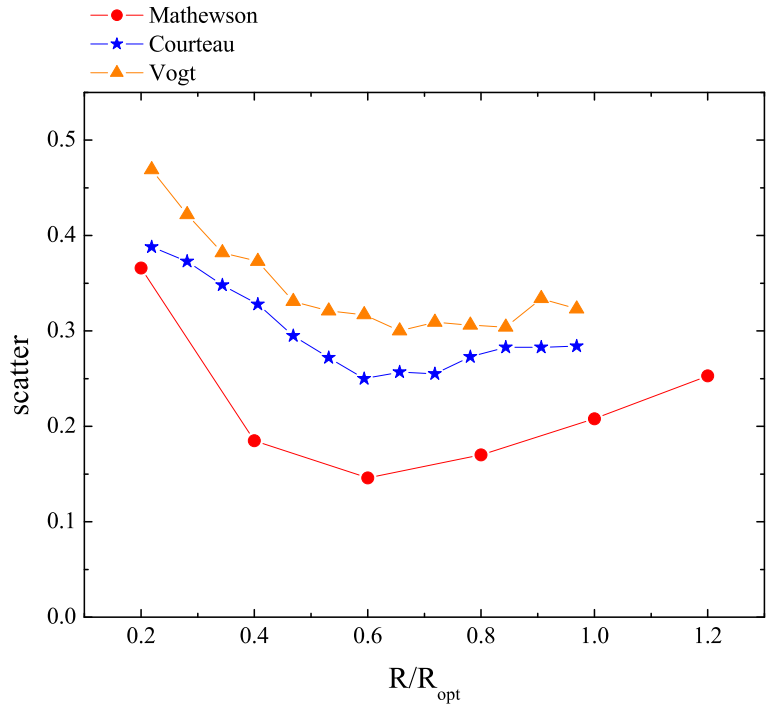}
%\epsfbox{fig1.eps}
\vspace{-1.cm} \caption{The scatter of the Radial Tully-Fisher
relation at different radii for the 3 samples.} \label{TF5}
\end{figure}
%%%%%%%%%%%%%%%%%%%%%%%%%%%%%%%%%%%fig5%%%%%%%%%%%%%%%%%%%%%%%%%%%%%%%%%%%

%%%%%%%%%%%%%%%%%%%%%%%%%%%%%%%%%%%fig6%%%%%%%%%%%%%%%%%%%%%%%%%%%%%%%%%%%
\begin{figure}\vspace{-5.cm}
\centering\epsfxsize=15.cm \epsfbox{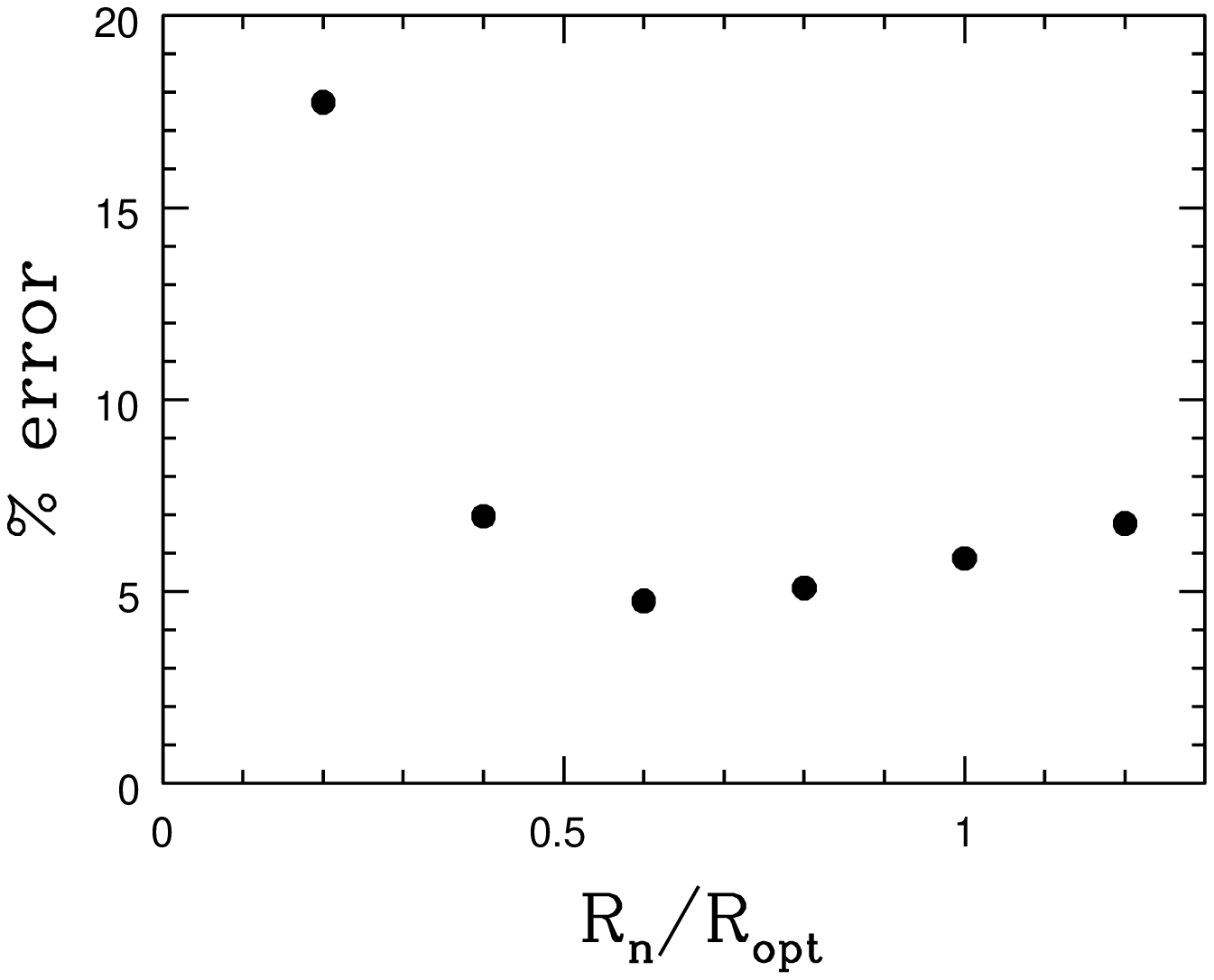}
%\epsfbox{fig1.eps}
%%\vspace{-1.cm}
\vspace{-1.cm}\caption{The $\%$ errors of the rotational velocity for different
radii} \label{TF6}
\end{figure}
%%%%%%%%%%%%%%%%%%%%%%%%%%%%%%%%%%%fig6%%%%%%%%%%%%%%%%%%%%%%%%%%%%%%%%%%%

It is worth to look at the scatter of the Radial TF relation (see
Fig.5). We find that, near the galactic center the scatter is large
$0.3 -0.4 $ dex, possibly due to a "random" bulge component
governing the local kinematics in this region being  almost
independent of the total galaxy magnitude. The scatter starts to
{\it decrease} with $R_n$, to reach a minimum of $0.15-0.3$ dex at
$R_n $ corresponding to two disk length scales, the radius where the
contribution of the disk to the circular velocity $V(R)$ reaches the
maximum. From $2 R_D$ onward, the scatter increases outward
reaching $0.3$ dex, at the farthest distances with available data,
i.e. at 3-4 disk length-scales.

%%%%%%%%%%%%%%%%%%%%%%%%%%%%%%%%%%%fig7%%%%%%%%%%%%%%%%%%%%%%%%%%%%%%%%%%%
\begin{figure}
\centering\epsfxsize=18.cm \epsfbox{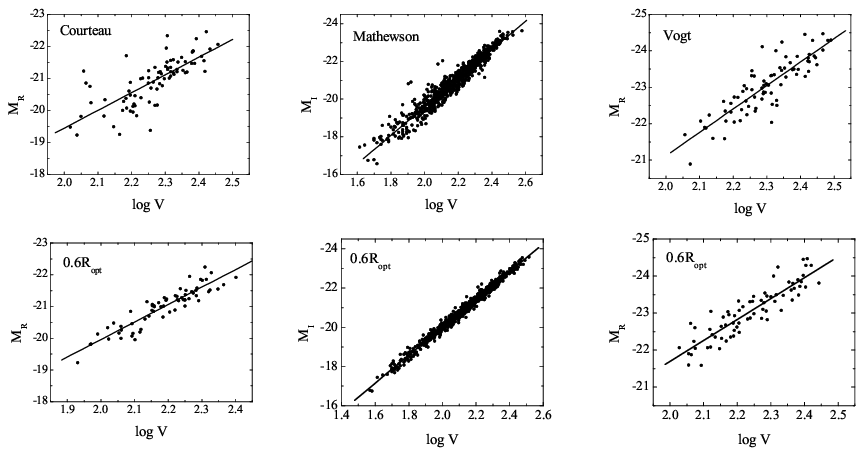}
%\epsfbox{fig1.eps}
\vspace{-1.cm} \caption{The standard TF relation for all 3 samples compered
with the RTF relation at $R=0.6R_{opt}$.}\label{TF7}
\end{figure}
%%%%%%%%%%%%%%%%%%%%%%%%%%%%%%%%%%%fig7%%%%%%%%%%%%%%%%%%%%%%%%%%%%%%%%%%%

Let us notice that  these  scatters are  remarkably small. Most of
the relations in the RTF family are statistically at least as
significant as the standard TF relation, while the most correlated
relationship, i.e. that at $2 R_D$ shows a  r.m.s of only 0.2 - 0.3
magnitudes (besides, significantly smaller than that of the standard
TF (see Figure 6)). Since the r.m.s include also the effects of
various observational errors, such  small values could  even
indicate that the intrinsic scatter  of the RTF relation at $\sim
2 R_D$ is almost zero.

An important consequence of the smallness of  the r.m.s of
TF-like relationships is that we can claim that  at {\it any
radius},  the luminosity {\it predicts} the rotation velocity
within the error of
$$
\delta V_n/V_n =  \ln(10) s_n/a_n\,. \eqno(10)
$$
This error (that includes also distance, inclination, and non
circular motions errors)  is  remarkably  small,  mostly lying
between 5\%-10\%, and not exceeding 20\%  even  in the very inner
bulge dominated regions (see figure for the PS95 sample). This
result is an additional proof for the URC paradigm, even more
impressive than the set of synthetic RC's in PSS.

Incidentally, the existence of a radius ($R=0.6R_{opt}$) at which
the TF-like relations  show  a minimum in the internal scatter is
not related to the overall capability of the luminosity in
predicting the rotation velocities. At very small $R \sim (0.1
-0.2)R_{opt}$ the (random) presence of a bulge increases the
scatter, in that, (as in ellipticals) the actual
kinematical-photometric fundamental plane likely includes a third
observational quantity (maybe the central surface density). At large 
$R$, the observed  small increase of the scatter can be avoided by
introducing small quadratic term in the eq (9).

The scatter of the RTF in the R band for the corresponding samples
is somewhat larger than that in the I band for   the PS95 sample.
This can be easily explained by the following:  i) the former
samples  include also a (small) fraction of Sa objects and
their RCs are of higher spatial resolution (lower bin size) and
therefore  less efficient in smoothing the non-circular motion
arisen by bars and spiral arms   ii) the R band is more affected
than the I Band  by  random recent episodes of star formation. A
conservative  estimate  of these effects is $\sigma_{obs}\geq  0.2$,
thus the intrinsic scatter of the RTF   in the R band
$(s_n^{2}-\sigma_{obs}^2)^{0.5}$ results similar to that found in 
the I band.

\section{Radial TF: implications}

The marked systematic increase of the slopes of  the RTF
relationship, as their reference radius $R_n$ increases  from
the galaxy center to the stellar disk edge, bears very important
consequences. First, it excludes, as viable mass models, those
in which:

  {\it i)} The gravitating mass follows the light distribution,
due  to a  total absence of non baryonic dark matter or to DM
being distributed similarly to the stellar matter. In both
cases, in fact, we do not expect to find any variation of the
slopes $a_n$ and very trivial variations of the zero points
$b_n$ with the reference radius $R_n $, contrary to the evidence
in Tables 1, 2, 3 and Fig. 4.

{\it ii)} The DM is present but with a  {\it luminosity
independent} fractional amount inside the optical radius. In
this case, in fact, the value of the circular velocity at any
reference radius $R_n$ will be a {\it luminosity independent}
fraction of the value at any other reference radius (i.e. $\log
V(R_n )=k_{nm} + \log V( R_m)$, $k_{nm}$ independent of
luminosity). As a consequence the slopes  $a_n $ in the RTF will
be independent of $R_n $, while the zero-points $b_n $ will
change in a characteristic way. It is clear  that the evidence
in Table 1, 2, 3 and Figure 4 contradicts this possibility.

The RTF contains crucial information on the mass distribution in
spirals.  In paper II we will fully recover and test it with
theoretical scenarios. Here, instead,   we will  use a simple mass
model (SMM),  that  includes a bulge, a disk and a halo mass
component and it is tunable by means of  4 free parameters. By
matching this model with the slopes of the RTF relation vs reference
radii relationship (hereafter SRTF) we derive the {\it gross }
features of the mass distribution in spirals. Let us point out, that
this method has a clear advantage with respect to
the mass modelling based on RCs. In this latter  procedure, since
the circular velocities have a quite limited  variation with radius,
physically different mass distributions   may reproduce the
observations equally well. Here instead, on one side, we will
use an observational quantity, the slope of the RTF relation, that
shows large variation with  reference radius; on the other side,
physically different mass distributions predict very different
slope vs reference radius relationships.

First, without any  loss of generality affecting our results, we
assume the well known relationships among the crucial structural
properties of spirals:  {\it i) }
$$
  R_D =R_1 l^{0.5}\,, \eqno(11a)
$$
see PSS, with $l\equiv 10^{(M_I-M_I^{max})/2.5}$ and
$M_I^{max}=-23.5 $ the maximum magnitude of our sample, and {\it
ii)}
$$
M_D =M_1 l^{1.3}\,, \eqno(11b)
$$
(e.g. Shankar et al. 2006 and references therein). Notice that the
constants $R_1$ and $M_1$ will play no role in the following. We
will best fit the $a_n$  data, i.e. the SRTF relation shown in Fig
(4) with the model function $a_{SMM} (x) $ derived from the SMM.
In this way we will fix the free structural mass parameters.

We describe in detail the adopted  SMM, the circular velocity is
a sum of three contributions
generated by the bulge component, taken as a point mass situated in
the center, a Freeman disk, and a dark halo.
$$
 V^2_{SMM}(x)= GM_D/R_D [f_d(x) + {M_b \over M_d} { 1 \over (3.2 x)}
 + {M_{halo}\over M_D}  { 1 \over 3.2} f_h(x, \alpha)],
$$
where $M_{halo}$ is the halo mass inside $R_{opt} $. It is useful
to measure $V^2_{SMM}$ in units of $G M_1/R_1$, and to set it to be
equal $1$.

The disk component  from  Eqs. (6) and (10),  (11) takes the form:

$$
   V_d^2(x,l) =  l^{0.8} f_d(x)\,,  \eqno(12)
$$
with  $V^2_{d}(1,1)=0.347$.

We set $M_b$ the bulge mass to be  equal to a fraction
$(c_b/(3.2\cdot 0.347) l^{0.5}$ of the disk mass, with $c_b$ a free
parameter of the  SMM  and the exponent $0.5$ is suggested by the
bulge-to disk vs total luminosity relation found for spirals. Then,
we get
$$
   V_{b}^2(x) =  c_b   V_d^2(1,l)  l^{0.5} x^{-1}\,. \eqno(13)
$$

The halo velocity contribution follows the profile of  Eq. 7; at
$ R_{opt}$ it is set to be equal to $  c_h/(3.2\cdot 0.347)
l^{(k_h - 0.5)} $ times the disk contribution $c_b$/$c_h$ and $ k_h $,
and $\alpha$  are the free parameters of the SMM.

%%$c_h/(3.2 0.347) $ indicates, at $(l,x)= (1,1) $ the
%%halo-to-disk   contribution to  the circular velocity, while
%%$k_h $  indicates  the value of the exponent of the
%%mass-to-light   power law relationship $M_{halo}(<R_{opt}) $ vs
%%$l^k_h $.

%%%%%%%%%%%%%%%%%%%%%%%%%%%%%%%%%%%fig8%%%%%%%%%%%%%%%%%%%%%%%%%%%%%%%%
\begin{figure}
\vspace{-1.cm}\centering\epsfxsize=14.cm \epsfbox{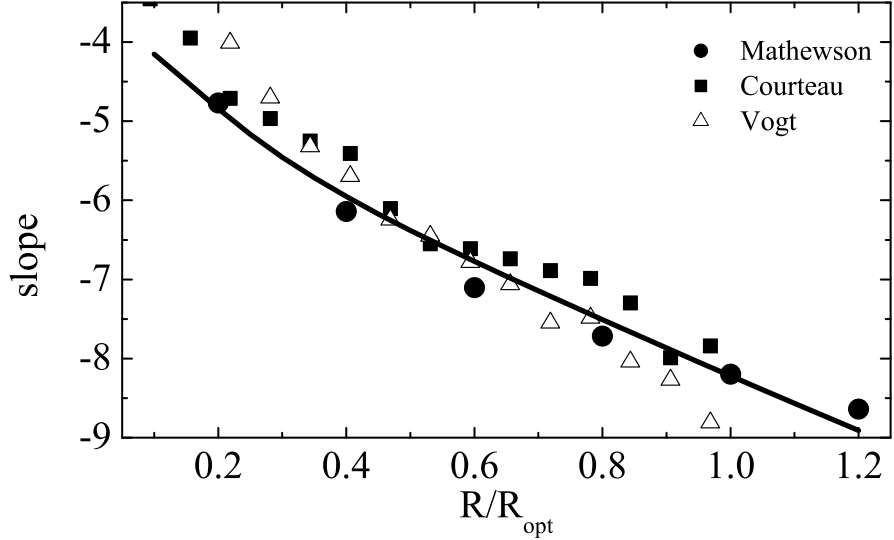}
%\epsfbox{fig1.eps}
\vspace{-1.cm} \caption{Slopes $a_n$ of the RTF for the 3 samples as
a function of the reference radius in units of $R_{opt}$. The solid line
is the best fit relation for Eq. 14}
\label{TF8}
\end{figure}
%%%%%%%%%%%%%%%%%%%%%%%%%%%%%%%%%%%fig8%%%%%%%%%%%%%%%%%%%%%%%%%%%%%%%%

Then, we can write:

$$
 V_{SMM}^2 (x,\alpha, l) = (c_b l^{1.3}/x + l^{0.8} f_d(x)  +
c_h l^{(k_h - 0.5)} f_h(x,\alpha))\,,          \eqno(14)
$$
where $f_d$ is given by Eq. (6),  $f_h$ by Eq. (7). Notice that
the simple form of $V_{SMM}$ allows us to get the predicted
slope parameter $a_{SMM}(x)$.

 The  core radius $\alpha $ is
a  DM free  parameter, however let us anticipate that, provided that
this quantity   lies between $0.5 $ and $2$ (see Donato et al. 2004)  it
does not  affect in a relevant way  the SMM predictions.  Instead
the $a_{SMM}(x)$ relationship strongly depends on
the values of  $c_h$, $k_h$, $c_b $,
and therefore they can be estimated with a good precision. We can
reproduce the observational $a_n=a(R_n)$ relationship,  by means of
the SMM (see Fig. 8) with the following best fit parameters values:
i) $k_h=0.79 \pm 0.04$, that means
that less luminous galaxies have larger fraction of dark matter, ii)
$c_b=0.13 \pm 0.03 $, and iii) $c_h= 0.13 \pm 0.06$ that indicates
that at $(l,x)=(1,1)$ (i.e. inside $R_{opt}$),
 $20\%$ of the mass is in the bulge component,
$20\%$ in the halo, while $60\%$ is in the stellar disk.  
The quoted uncertainties are
the formal $\chi^2 $ fitting uncertainties.

%%%%%%%%%%%%%%%%%%%%%%%%%%%%%%%%%%%fig9%%%%%%%%%%%%%%%%%%%%%%%%%%%%%%%%%%%
\begin{figure}
\vspace{-1.cm}\centering\epsfxsize=12.cm \epsfbox{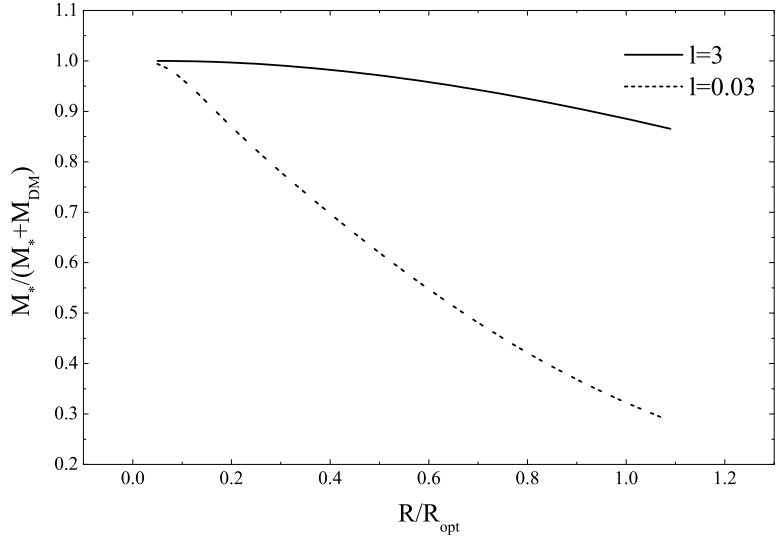}
%\epsfbox{fig1.eps}
\vspace{-1.cm} \caption{Baryonic mass fraction as a function of
normalized radius for high and low luminosity objects.}
\label{TF9}
\end{figure}
%%%%%%%%%%%%%%%%%%%%%%%%%%%%%%%%%%%fig9%%%%%%%%%%%%%%%%%%%%%%%%%%%%%%%%%%%

%%%%%%%%%%%%%%%%%%%%%%%%%%%%%%%%%%%fig10%%%%%%%%%%%%%%%%%%%%%%%%%%%%%%%%%%%
\begin{figure}
\vspace{-1.cm}\centering\epsfxsize=12.cm \epsfbox{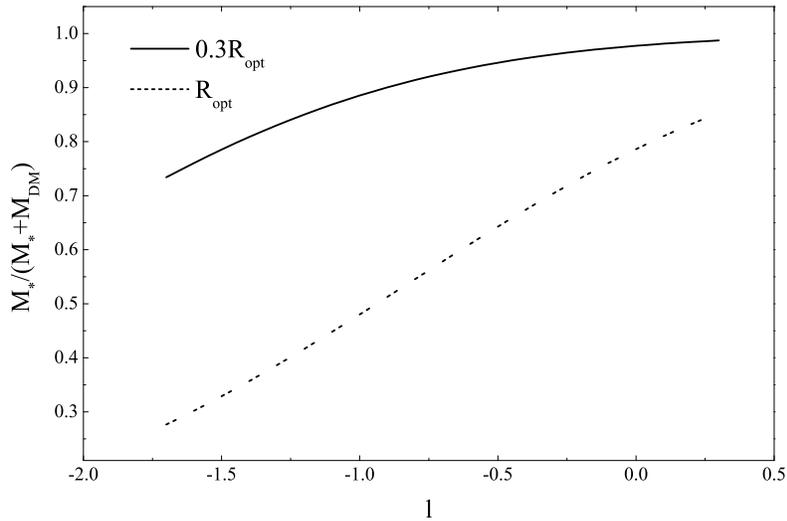}
%\epsfbox{fig1.eps}
\vspace{-1.cm} \caption{Baryonic mass fraction at two different
radii as a function of luminosity.} \label{TF10}
\end{figure}
%%%%%%%%%%%%%%%%%%%%%%%%%%%%%%%%%%%fig10%%%%%%%%%%%%%%%%%%%%%%%%%%%%%%%%%%%

\section{Discussion and conclusions}

In spirals,  at different galactocentric distances measured in
units of disk length-scales $j R_D$ ($j=0.2, ..., 4$), there
exists a family of independent Tully-Fisher-like relationships,
$M_{band} = b_j + a_j \log V (R_j)$, that we call the Radial
Tully-Fisher relation, that contains crucial information on
the mass distribution in these objects. In fact:

1)  The RTF relationships show large systematic variations in
 their  slopes $a_j$ (between $-4$ and $-8$)  and a r.m.s.
 scatter  generally  smaller  than that of the standard TF. This
 rules out the case in which the light follows the gravitating
 mass, and, in particular, all  mass  models that imply:  a) an
 absence of  dark matter,  b) a single mass component c)  the same
 dark-to-luminous-mass fraction within $j R_D$ in all galaxies.

2)  The slopes $a_j$  decrease monotonically with $R_j $, implies
the presence of a non luminous mass  component whose dynamical
importance,  with respect to the stellar disk, increases with radius.

3)  The existence of the RTF and the features of the slope vs $R_j $
can be well reproduced by means of a three components mass model that
includes: a dark halo, with a core of radius $4 R_D$ and mass
$M_{halo}(R_{opt}) \propto  l^{0.7} $, a central bulge with
$M_{bulge} \propto l^{1.8} $, an exponential thin disk of mass
$M_d \propto l^{1.3}$ with, at $10^{11} L_{B,\odot}$, $60\%$ of
the mass inside $R_{opt}$ in stellar form.

Let us also notice that we have  produced a qualitatively  new
evidence for the presence of a luminosity dependent  mass
discrepancy in spirals, different from that obtained from the non
Keplerian shapes of the RC's. While, the latter originates  from a
failure: we {\it do not} observe the Keplerian fall-off of the
circular velocities at the disk edge, and therefore we must
postulate a new component, here, we provide a  { \it positive}
evidence for the existence of such dark component: we {\it detect}
radial change of the slope and the scatter of existing relations
between observables  that positively indicates the presence of a
more diffuse dark component.

The small scatter of the  TF-like relationships in the inner regions
of the spirals ($  x=0.4-0.6  $) has important implications
for the DM halo core vs cusp controversy, more specifically on the claim
that the observed rotation velocities $ V(R)$ do not coincide
with  the circular velocities, i.e. with the equilibrium velocities
associated with  the central galaxy gravitational  potential $\Phi$,
$V_{circ} (R)=  (-R d\Phi/dR)^{1/2})$.  It is claimed that, due
non-circular/non-central motions, gas pressure gradients and other
effects,  $ V(R) <  V_{circ} $ by a substantial amount and
models in which observed  cored-solid body  rotation curves $V(R)$ underly
actual NFW-cusped  circular velocities $V_{circ}(R)= V_{NFW}(R) \neq V(R)$
have been built (Hayashi 2006). Let us notice
that this effect must be very large in that it must explain the
large discrepancy between the observed rotation curves (taken as
bona-fide circular velocities) and their NFW best fit mass models:
the typical difference is
$$
(V  -  V_{NFW})= (20 - 60) km/s
$$
for galaxies with $V_{opt}=100-150$km/s (Gentile et al., Donato et
al. and references therein).  At $R_n= (0.4 - 0.6) R_{opt} $   we have
$s_n \sim  0.3$,  that  certainly includes a component due to
distance and inclinations errors. Even if  we assume that its all
due to the fact that $V \neq V_{circ}$ this would trigger a discrepancy 
of only $ \sim 5 - 15$ km/s, far  too small to account for the severe
discrepancies of the NFW models.

Finally, let us stress that any model of  formation of spiral
galaxies must be able to produce   (e.g. in the I band) a $M_I$
vs $\log V(2R_D)$  relationship with a slope of $7\pm 0.1$ and
an intrinsic scatter of $\leq 0.15$ magnitudes.

\subsection*{Acknowledgments}
We thank S. Courteau and N. Vogt for kindly provided data. We also
thank Gianfranco Gentile for helping us with preparation of this
paper. We want to thank the anonymous referee for detailed comments
that much improved the final version of this paper.

\vfill\eject

\vfill\eject

\null \vspace{0.4truecm}

\section{Appendix}
In this appendix we present tables and figures related to our result
and the galaxies of Samples 2 and 3.
\bigskip

\begin{table}[ht]
\caption{Parameters of the Radial Tully-Fisher relation at
different radii for the Mathewson sample} \label{tbl1}
\begin{tabular}{llllllll lr}
$R/R_{opt}$    & zero point & error &slope &error &SD &N & \\
               &     &   \\
0.2&  -11.78&0.103&-4.77&0.054&0.366&739&\\
0.4&  -8.241&0.068&-6.141&0,033&0.185&786&\\
0.6& -5.787&0.063&-7.102&0.029&0.146&794&\\
0.8&  -4.22&0.09&-7.718&0.042&0.17&657&\\
1.0&  -3.034&0.146&-8.197&0.067&0.208&447&\\
1.2& -1.979&0.261&-8.639&0.118&0.253&226&\\
 \vspace{0.15cm}
\end{tabular}

column 1 - the isophotal radius,\\ column 2 - intercept value
$b_n$,\\ column 3 - the standard error of $b_n$, \\ column 4 -
the slope $a_n$,\\ column 5 - the standard error of $a_n$,\\
column 6 - the scatter,\\ columns 7 - the number of
observational points
\end{table}

\begin{table}[ht]
\caption{Parameters of the Radial TF relation at different radii for the
Courteau sample} \label{tbl2}
\begin{tabular}{llllllll lr}
$R/R_{opt}$    & zero point&error&slope&error&SD&N& \\
               &     &   \\
0.03&-18.217&0.287&-1.8&0.189&0.516&75&\\
0.09&-15.615&0.39&-2.878&0.209&0.381&74&\\
0.16&-14.355&0.495&-3.336&0.249&0.388&75&\\
0.22&-13.379&0.55&-3.707&0.27&0.397&76&\\
0.28&-12.155&0.741&-4.194&0.355&0.465&79&\\
0.34&-11.7&0.613&-4.374&0.291&0.361&74&\\
0.41&-11.11&0.617&-4.586&0.289&0.338&72&\\
0.47&-9.962&0.674&-5.086&0.31&0.295&68&\\
0.53&-9.114&0.699&-5.445&0.321&0.296&71&\\
0.59&-8.919&0.733&-5.519&0.334&0.293&68&\\
0.66&-8.623&0.752&-5.61&0.341&0.279&65&\\
0.72&-8.351&0.625&-5.737&0.284&0.255&63&\\
0.78&-8.172&0.763&-5.799&0.345&0.274&61&\\
0.84&-7.329&1.0&-6.152&0.456&0.32&53&\\
0.91&-6.372&1.0&-6.58&0.449&0.286&44&\\
0.97&-7.573&1.54&-6.042&0.688&0.31&37&\\
1.03&-7.728&1.993&-5.984&0.888&0.311&25&\\
1.09&-9.265&2.254&-5.307&1.0&0.212&14&\\
1.16&-6.853&1.767&-6.35&0.789&0.281&15&\\
 \vspace{0.15cm}
\end{tabular}
\end{table}

\begin{table}
%%\vspace{-1.cm} %%[ht]
\caption{Parameters of the Radial Tully-Fisher relation at
different radii for the Vogt sample} \label{tbl3}
\begin{tabular}{llllllll lr}
$R/R_{opt}$    & zero point&error&slope&error&SD&N& \\
               &     &   \\
0.09&-19.351&0.526&-1.992&0.278&0.55&78&\\
0.16&-18.118&0.64&-2.456&0.316&0.528&78&\\
0.22&-16.142&0.714&-3.309&0.339&0.472&76&\\
0.28&-14.787&0.728&-3.869&0.338&0.43&77&\\
0.34&-13.583&0.73&-4.365&0.334&0.394&77&\\
0.41&-12.646&0.78&-4.747&0.354&0.386&77&\\
0.47&-11.746&0.784&-5.112&0.353&0.365&77&\\
0.53&-11.342&0.805&-5.264&0.361&0.361&75&\\
0.59&-10.698&0.778&-5.51&0.347&0.327&72&\\
0.66&-9.804&0.77&-5.885&0.341&0.309&71&\\
0.72&-9.244&0.83&-6.125&0.368&0.318&70&\\
0.78&-9.227&0.936&-6.104&0.414&0.341&68&\\
0.84&-7.873&0.906&-6.7&0.4&0.304&60&\\
0.91&-7.435&1.057&-6.893&0.466&0.334&58&\\
0.97&-6.377&1.153&-7.343&0.505&0.323&50&\\
1.03&-8.398&1.414&-6.435&0.62&0.345&41&\\
1.09&-7.953&1.377&-6.628&0.599&0.301&35&\\
1.16&-8.683&1.38&-6.947&0.622&0.228&24&\\
1.22&-9.616&1.444&-6.842&0.714&0.275&23&\\
1.28&-7.394&2.467&-6.834&1.072&0.347&16&\\
 \vspace{0.15cm}
\end{tabular}
\end{table}

\medskip

\begin{table}
%%\vspace{-5.cm}
\caption{Parameters of the standard Tully-Fisher relation for 3
samples.}\label{tbl4}
\begin{tabular}{lllllll lr}
Data    &zero point &error &slope &error &SD &N & \\
               &     &   \\
Mathewson&-4.455&0.15&-7.57&0.069&0.327&841&\\
Courteau&-8.398&1.26&-5.526&0.556&0.495&81&\\
Vogt et al.&-8.277&0.997&-6.423&0.433&0.389&79&\\
 \vspace{0.15cm}
\end{tabular}
\end{table}

%%%%%%%%%%%%%%%%%%%%%%%%%%%%%fig11%%%%%%%%%%%%%%%%%%%%%%%%%%%%%%%

\begin{figure}
\centering\epsfxsize=15.cm \epsfbox{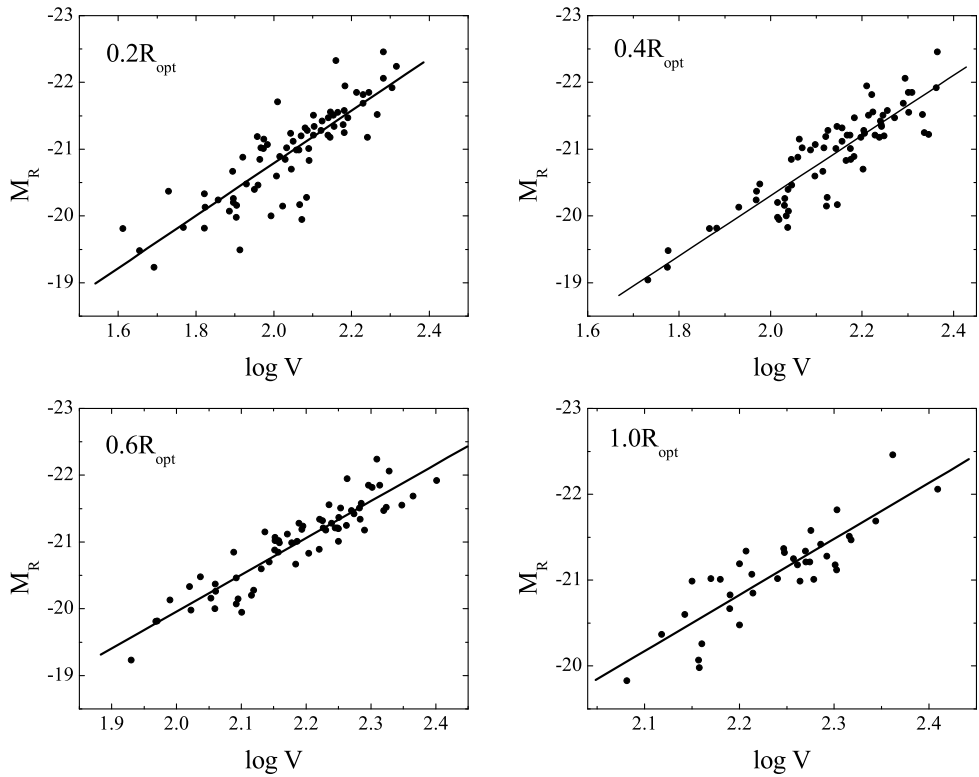}
%\epsfbox{fig1.eps}
\vspace{-1.cm} \caption{The Radial Tully-Fisher relation for the
Courteau sample.} \label{c-all}
\end{figure}

%%%%%%%%%%%%%%%%%%%%%%%%%%%%%%%fig11%%%%%%%%%%%%%%%%%%%%%%%%%%%%%%

%%%%%%%%%%%%%%%%%%%%%%%%%%%%%%%fig12%%%%%%%%%%%%%%%%%%%%%%%%%%%%%%%

\begin{figure}
\vspace{-1.cm}
\centering\epsfxsize=15.cm \epsfbox{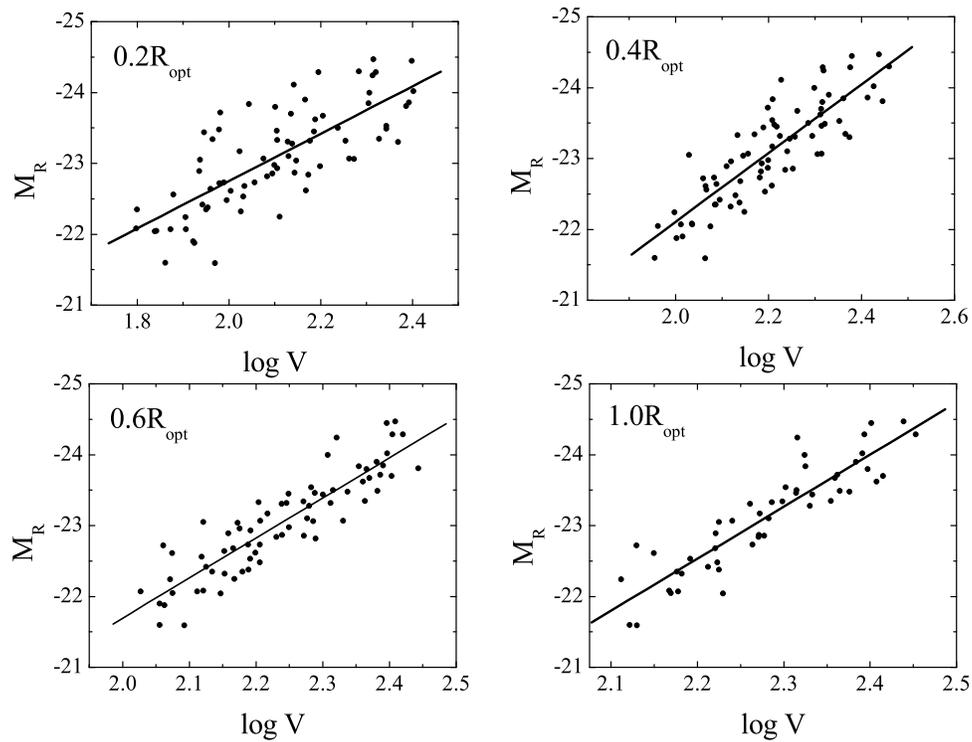}
%\epsfbox{fig1.eps}
\vspace{-1.cm} \caption{The Radial Tully-Fisher relation for the
Vogt sample. } \label{V-all}
\end{figure}

%%%%%%%%%%%%%%%%%%%%%%%%%%%%%%%%%%fig12%%%%%%%%%%%%%%%%%%%%%%%

\begin{table}
Names of galaxies from the Courteau sample: UGC 10096, UGC
10196, UGC 10210, UGC 10224, UGC 1053, UGC 10545, UGC 10560, UGC
10655, UGC 10706, UGC 10721, UGC 10815, UGC 11085, UGC 11373,
UGC 1152, UGC 11810, UGC 12122, UGC 12172, UGC 12200, UGC 12294,
UGC 12296, UGC 12304, UGC 12325, UGC 12354, UGC 12598, UGC
12666, UGC 1426, UGC 1437, UGC 1531, UGC 1536, UGC 1706, UGC
1812, UGC 195, UGC 2185, UGC 2223, UGC 2405, UGC 2628, UGC 3049,
UGC 3103, UGC 3248, UGC 3269, UGC 3270, UGC 3291, UGC 3410, UGC
346, UGC 3652, UGC 3741, UGC 3834, UGC 3944, UGC 4232, UGC 4299,
UGC 4326, UGC 4419, UGC 4580, UGC 4779, UGC 4996, UGC 5102, UGC
540, UGC 562, UGC 565, UGC 5995, UGC 6544, UGC 6692, UGC 673,
UGC 7082, UGC 732, UGC 7549, UGC 7749, UGC 7810, UGC 7823, UGC
783, UGC 784, UGC 8054, UGC 8118, UGC 8707, UGC 8749, UGC 8809,
UGC 890, UGC 9019, UGC 9366, UGC 9479, UGC 9598, UGC 9745, UGC
9753, UGC 9866, UGC 9973
\end{table}

\vspace{1.0cm}

\begin{table}
Names of galaxies from the Vogt sample: UGC 927, UGC 944, UGC
1033, UGC 1094, UGC 1437, UGC 1456,UGC 1459, UGC 2405, UGC 2414,
UGC 2426, UGC 2518, UGC 2618, UGC 2640, UGC 2655, UGC 2659, UGC
2700, UGC 3236, UGC 3270, UGC 3279, UGC 3289, UGC 3291, UGC
3783, UGC 4275, UGC 4287, UGC 4324, UGC 4607, UGC 4655, UGC
4895, UGC 4941, UGC 5166, UGC 5656, UGC 6246, UGC 6437, UGC
6551, UGC 6556, UGC 6559, UGC 6718, UGC 6911, UGC 7845, UGC
8004, UGC 8013, UGC 8017, UGC 8108, UGC 8118, UGC 8140, UGC
8220, UGC 8244, UGC 8460, UGC 8705, UGC 10190, UGC 10195, UGC
10459, UGC 10469, UGC 10485, UGC 10550, UGC 10981, UGC 11455,
UGC 11579, UGC 12678, UGC 12755, UGC 12792, UGC 150059, UGC
180598, UGC 210529, UGC 210559, UGC 210629, UGC 210634, UGC
210643, UGC 210789, UGC 211029, UGC 220864, UGC 221174, UGC
221206, UGC 251400, UGC 260640, UGC 260659, UGC 330781, UGC
330923, UGC 330925, UGC 330996, UGC 331021
\end{table}

\end{document}